\documentclass[ aip, apl, amsmath,amssymb, reprint] {revtex4-1}

\usepackage{amsmath, graphicx}

\begin{document}

\title{Plasma etching of superconducting Niobium tips for scanning tunneling microscopy}

\author{A. Roychowdhury}
\affiliation{Laboratory for Physical Sciences, College Park, MD 20740}
\affiliation{Center for Nanophysics and Advanced Materials, Dept. of Physics, University of Maryland, College Park, MD 20742}
\author{R. Dana}
\affiliation{Laboratory for Physical Sciences, College Park, MD 20740}
\author{M. Dreyer}
\affiliation{Laboratory for Physical Sciences, College Park, MD 20740}
\author{J. R. Anderson}
\affiliation{Center for Nanophysics and Advanced Materials, Dept. of Physics, University of Maryland, College Park, MD 20742}
\author{C. J. Lobb}
\affiliation{Center for Nanophysics and Advanced Materials, Dept. of Physics, University of Maryland, College Park, MD 20742}
\author{F. C. Wellstood}
\affiliation{Center for Nanophysics and Advanced Materials, Dept. of Physics, University of Maryland, College Park, MD 20742}
\date{\today}

\begin{abstract}
We report a reproducible technique for the fabrication of sharp superconducting Nb tips for scanning tunneling microscopy (STM) and scanning tunneling spectroscopy. Sections of Nb wire with 250 $\mu$m diameter are dry etched in an SF$_6$ plasma in a Reactive Ion Etcher. The gas pressure, etching time and applied power are chosen to produce a self-sharpening effect to obtain the desired tip shape. The resulting tips are atomically sharp, with radii of less than 100 nm, and generate good STM images and spectroscopy on single crystal samples of Au(111), Au(100), and Nb(100), as well as a doped topological insulator Bi$_2$Se$_3$ at temperatures ranging from 30 mK to 9 K.
\end{abstract}

\maketitle

\indent\indent Scanning tunneling microscopy (STM) and scanning tunneling spectroscopy are powerful techniques for probing the local density of states of conducting materials on the atomic scale.\cite{Binnig82} STM tips are typically fabricated out of normal metals such as Pt-Ir or W that have an approximately constant density of states (DOS) near the Fermi level.\cite{Musselman90, Ibe90, Oliva96} The tunneling conductance $\text{d}I/\text{d}V$ versus bias voltage $V$, between such a tip and sample, provides a direct measurement of the  local DOS of the sample. However, a normal metal tip is subject to Fermi broadening. By using a superconducting STM tip rather than a normal metal tip, one can obtain enhanced spectroscopic resolution due to the singularity at the gap edge in the superconducting DOS. \cite{usmakov01, Rodrigo04, Guillamon08, Noat10} Enhanced resolution is important in observing many phenomena, including superconducting gap anisotropy and multiband superconductivity.\cite{Noat10} Superconducting STM tips can also be used to directly probe the superconducting condensate on the atomic scale. \cite{Rodrigo06, Kohen05, Bergeal08, Sullivan13} Furthermore, superconducting STM tips could be used to probe the surface state of topological insulators such as Bi$_\text{2}$Se$_\text{3}$, and to explore superconductor/topological insulator interfaces, expected to support Majorana bound states.\cite{Fu08, Hasan10, Stanescu10, Linder10} In this Letter we describe the fabrication of superconducting Nb STM tips  using a reactive ion etcher and demonstrate their performance.

Despite their potential advantages, it has proven challenging in practice to reproducibly fabricate superconducting tips that are mechanically robust and sharp enough to obtain atomic resolution. Common superconducting elements such as Al and Pb  oxidize quickly in air. Although significant progress has been made in the development of Nb tips using a variety of techniques, \cite{Pan98, Kohen05, Suderow02, Rodrigo04, Ternes06} they are difficult to reproduce, and require UHV tip exchange capabilities or a coarse x-y stage, approaches that are difficult to implement in a scanning probe microscope that is mounted on a dilution refrigerator. Superconducting tips based on the proximity effect \cite{Naaman01, Kimura08} render a UHV environment unnecessary, but it appears this approach has yet to yield atomic resolution topographic images. Although materials such as MgB$_2$ and high-$T_\text{c}$ superconductors are brittle and difficult to work with due to surface degradation, small crystals of high $T_\text{c}$ superconductors have been successfully used as STM tips by attaching them to the end of a Pt-Ir wire.\cite{Giubileo01, Xu03, Bergeal08, Noat10} While this method has yielded stable superconducting tips with good spatial resolution, the anisotropic wave function of these superconductors renders spectroscopic information about the sample more complicated to deconvolve than an s-wave BCS density of states.
 
\begin{figure}[t]
\includegraphics[width=0.45 \textwidth]{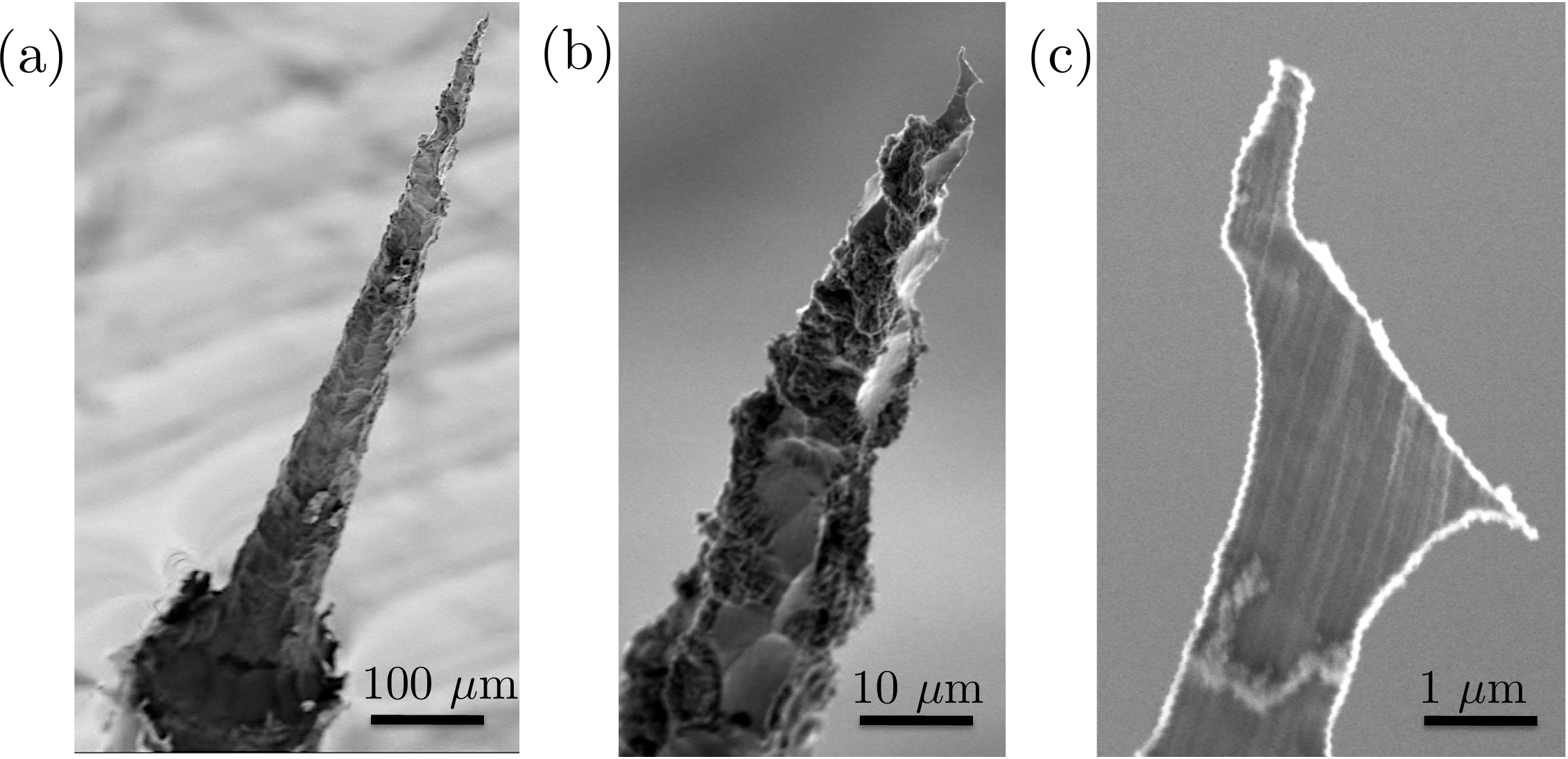}
\caption{Scanning electron microscope images of a single Nb tip fabricated in a reactive ion etcher at length scales (a) 100 $\mu$m, (b) 10 $\mu$m and (c) 1 $\mu$m.}
\end{figure}

We chose to make our tips from Nb for several reasons. Nb is a conventional $s$-wave superconductor with a transition temperature $T_\text{c}$ of 9.3 K. This allows us to operate up to a relatively high cryogenic temperature. Also, Nb has slower growing oxides than Al which allows for {\it ex situ} fabrication. Furthermore, Nb is very mechanically robust, much more so than Pb, Sn, or In for example, which allows for better imaging. Finally, techniques for patterning thin film and bulk Nb are well known from the fabrication of superconducting microelectronics circuits and rf cavities. \cite{Lichtenberger93,Numata01, Zhong12} 

To etch the tips, we place 50 mm long pieces of bare 250 $\mu$m diameter Nb wire vertically in an SF$_6$ plasma generated by a reactive ion etcher. \footnote{Plasma-Therm 790 series: http://www.plasmatherm.com/790-rie.html} Fifteen tips are etched at a time, held in holes drilled in a 12 cm $\times$ 12 cm $\times$ 2 cm Al block. The tips are glued into the holes with photoresist that is cured for 10-15 minutes at 150$^\circ$ C. We found that this prevents the formation of split, double, or triple tips. The recipe we converged on uses an SF$_6$ flow rate of 10 $\text{cm}^3$/s, a pressure of 100 mTorr, and an applied rf power of 150 W. The resulting dc self-bias potential of the cathode is typically around 52 V. The typical etch time is about 90 minutes. 

Figure 1(a) shows an SEM image of a tip after etching. All of the exposed section of the wire has been etched away except for a 600-800 $\mu$m long tapered conical apex with an angle of about 8$^\circ$ [see Fig.~1(a)]. The base of the tapered region is defined by the photoresist and Al holder. The aspect ratio of the cone was controlled primarily by the plasma gas pressure, which determines the ratio of isotropic to anisotropic etch rates.\cite{Madou02} Pressures lower than 100 mTorr resulted in more anisotropic etching due to longer mean free paths of the ions. The precise geometry of the tip apex varied somewhat from tip to tip, and we chose tips that appeared to be sharpest and mechanically sound. Completed tips were transported in air and mounted on an STM. The STM chamber was then evacuated within 60-90 minutes to limit oxide growth.

\begin {figure}[t]
\centering
\includegraphics[width= 0.5\textwidth]{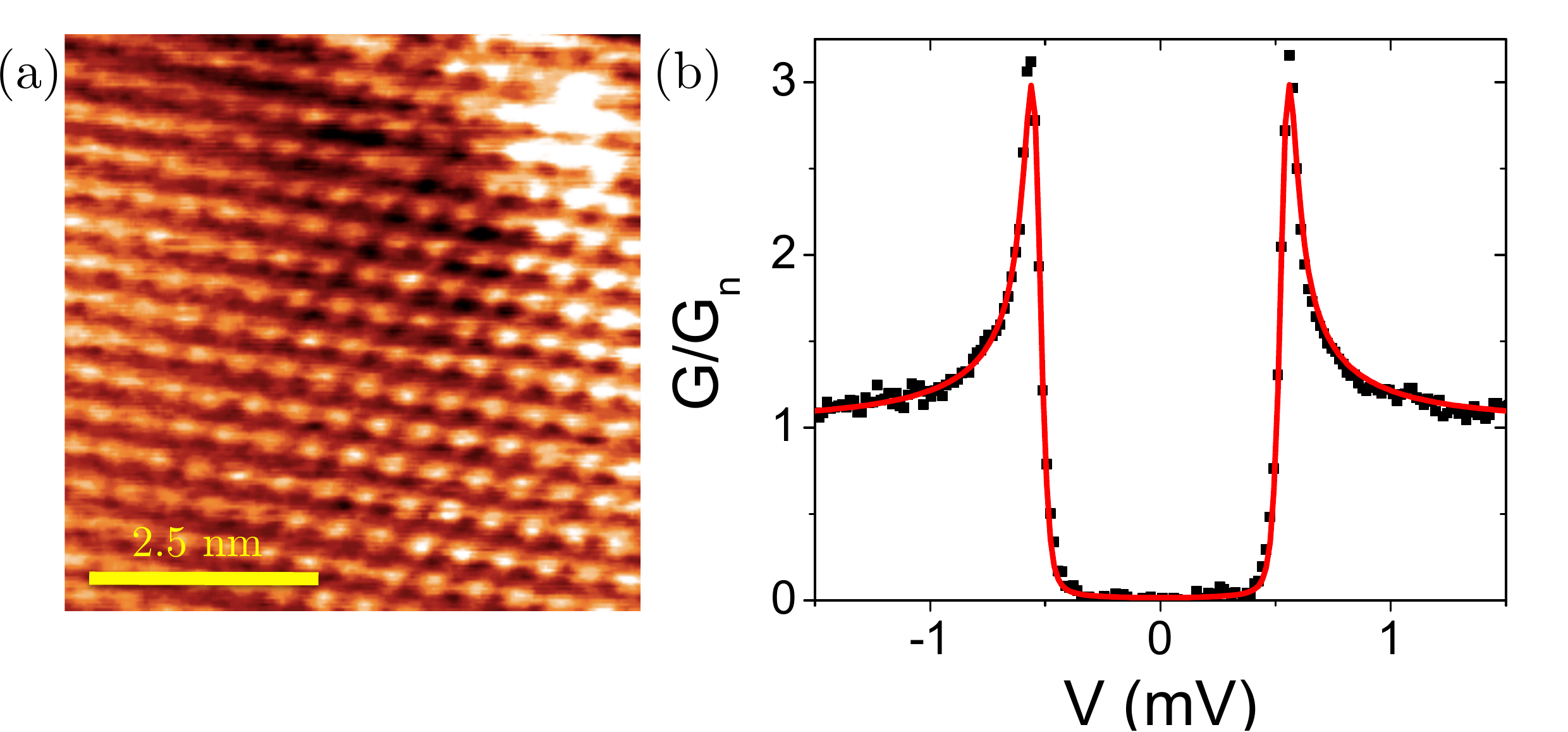}
\caption{(a) Atomic resolution image taken with a Nb tip on a Bi$_\text{2}$Se$_\text{3}$ sample at 35 mK. (b) Plot of normalized conductance G/G$_\text{n}$ vs. tip-to-sample voltage $V$ for the Nb tip and  Bi$_\text{2}$Se$_\text{3}$ sample shown in (a). Black points are measured data and red curve is fit to Eq.~1 with energy gap $\Delta = 0.54$ meV, effective temperature $T_\text{eff}=184$ mK, and broadening $\Gamma \approx 10^{-5}$ meV.}
\end{figure}

\begin {figure}[t]
\centering
\includegraphics[width= 0.49\textwidth]{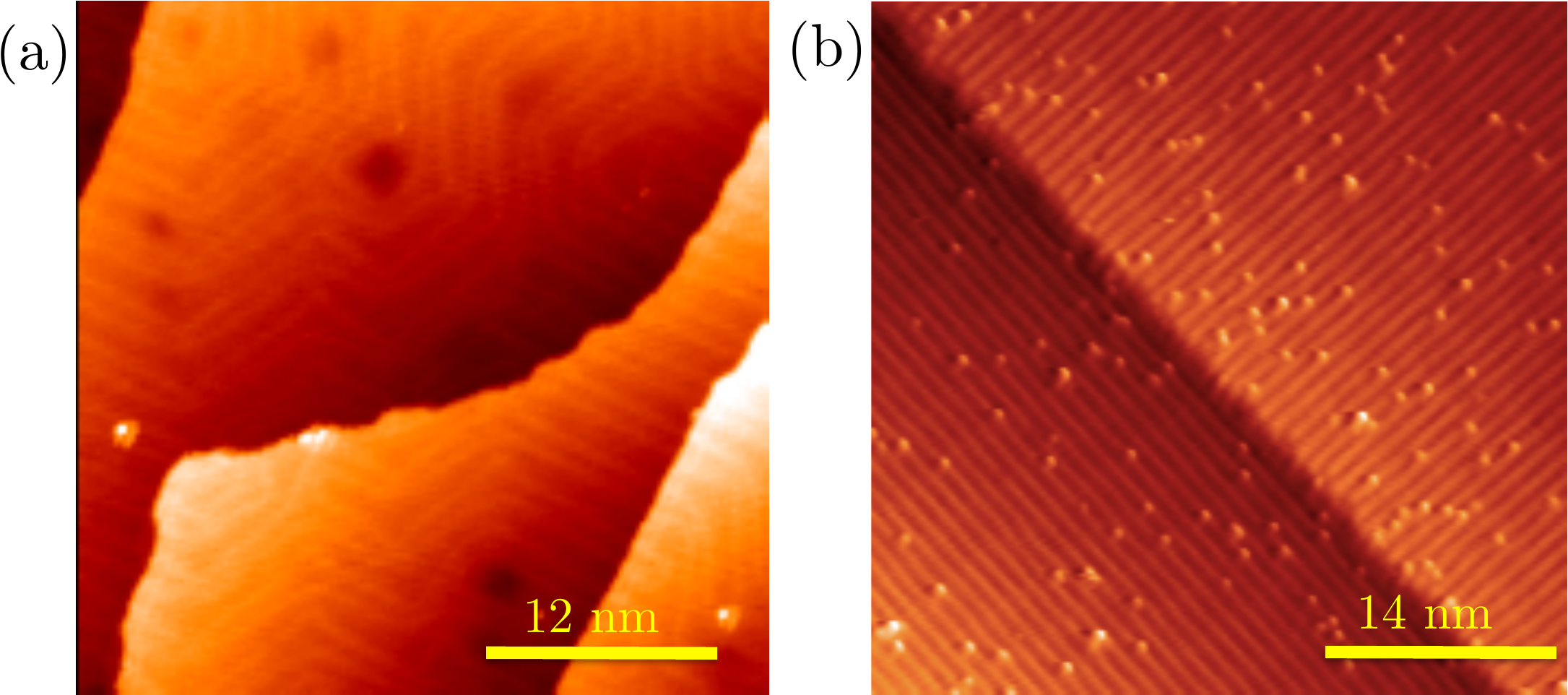}
\caption{(a) Topographic images of  Au(111) taken with a Nb tip at 4.2 K shows Herringbone reconstruction,\cite{Barth90, Narasimhan92} and (b) Au(100) sample at 1.5 K  showing a striped reconstruction.\cite{Binnig84}}
\end{figure}

Figures 1(b) and 1(c) show, at the 10 $\mu$m scale and at the 1 $\mu$m scale, respectively, that the surface of the etched tips is rough at these scales. Typically our etch conditions result in a fairly isotropic etching process, except for geometry dependent variations in the local electric field and sheath properties. The sharp spiked microstructures seen in Figs.~1(b) and 1(c) are evidence for self sharpening which can occur when the Child-Langmuir (steady state) sheath width is much larger than the radius of curvature of the structure being etched.\cite{Watterson89} For our parameters, the expected sheath width that forms conformally around the wire is typically 0.1-1 mm, which is indeed much larger than the 10-100 nm tip radius. Since nearly all of the potential is dropped across this sheath, its structure around the cathode significantly affects the ion trajectories and impact angles, resulting in geometry dependent etch rates that vary locally. Analysis of complex cathode geometries suggest that the ion impact rate and angle in the vicinity of protrusions on the surface of the wire create self sharpening effects.\cite{Donnelly89, Watterson89, Sheridan09} 

Although the fine-scale geometry of our Nb tips is unconventional for STM probes, the tips have yielded good topographic and spectroscopic data. In our STM setup we typically use field emission to clean the tip. This is done by turning off the feedback loop controlling the tip-to-sample separation, retracting the tip a few tens of nms from the sample surface, and ramping the tip-to-sample voltage up to 80-100 V. The voltage is ramped up until the tip-to-sample current drops to zero, indicating that a piece of the tip has fallen off. The roughness of the etched Nb surface increases the chance of regaining a sharp tip apex, thus extending the usability of our tip. In practice we have run up to 30 field emissions on a single Nb tip, over a period of around 6 months. 

To test the resolution and stability of our tips, we used them in two cryogenic STMs to examine single crystals of Au(111), Au(100), Nb(100), and the topological insulator Bi$_2$Se$_3$.  Figure 2(a) shows an atomically resolved, unfiltered topographic image of the Bi$_2$Se$_3$ sample, indicating that the Nb tip was both mechanically stable and atomically sharp. Figure 2(b) shows a measured $\text{d}I/\text{d}V$ curve on this sample. This data was taken at 30 mK  using a millikelvin STM.\cite{Roychowdhury13} The red curve in Fig.~2(b) shows a fit to 
\begin{equation}
\frac{dI}{dV} = G_\text{n} \int_{-\infty}^{\infty} g(E+eV) N_\text{s,tip}(E)dE
\end{equation}
where $G_\text{n}$ is the normal state conductance,  $g(E+eV)= -\partial f(E+eV)/\partial E$, $f(E)$ is the Fermi function at energy $E$, and $N_\text{s,tip}(E)$ is the normalized local density of states of the tip. For a superconducting tip with a BCS density of states and a finite quasiparticle relaxation rate $\Gamma$ \cite{Dynes78}, 
\begin{equation}
N_\text{s,tip}(E) = \text{Re} \left[ \frac{ \vert E- i\Gamma \vert}{\sqrt{(E - i\Gamma)^2 - \Delta^2}}\right] .
\end{equation}
For the fit in Fig.\ 2(b) we find $\Delta = 0.54$ meV and $\Gamma \ll \Delta$. Our tips typically exhibit gap values that vary from 0.54 -- 1.4 meV, comparable to the variation seen in other Nb STM tips.{\cite{Pan98, Ternes06} We attribute this variation to finite size effects and local variations in the gap due to the diffusion of oxygen into the atomically sharp tip apex.

Figure 3 shows topographic images of atomic reconstructions on surfaces of Au(111) and Au(100) single crystals scanned at 4.2 K and 1.5 K respectively. Figure 3(a) was taken on a 4 K STM system,\cite{Dreyer10} while Fig.\ 3(b) was obtained using our millikelvin STM.\cite{Roychowdhury13} A series of d$I$/d$V$ curves measured on this sample between 1.5 K and 9 K are shown in Fig.~4. The temperature was varied by applying power to the mixing chamber of the dilution refrigerator, while extracting the $^3$He-$^4$He mixture from it. A RuO$_x$ thermometer at the mixing chamber was used to record the temperature of the measurements. The evolution of the superconducting gap with temperature is consistent with BCS theory. 

\begin {figure}[t]
\centering
\includegraphics[width= 0.49\textwidth]{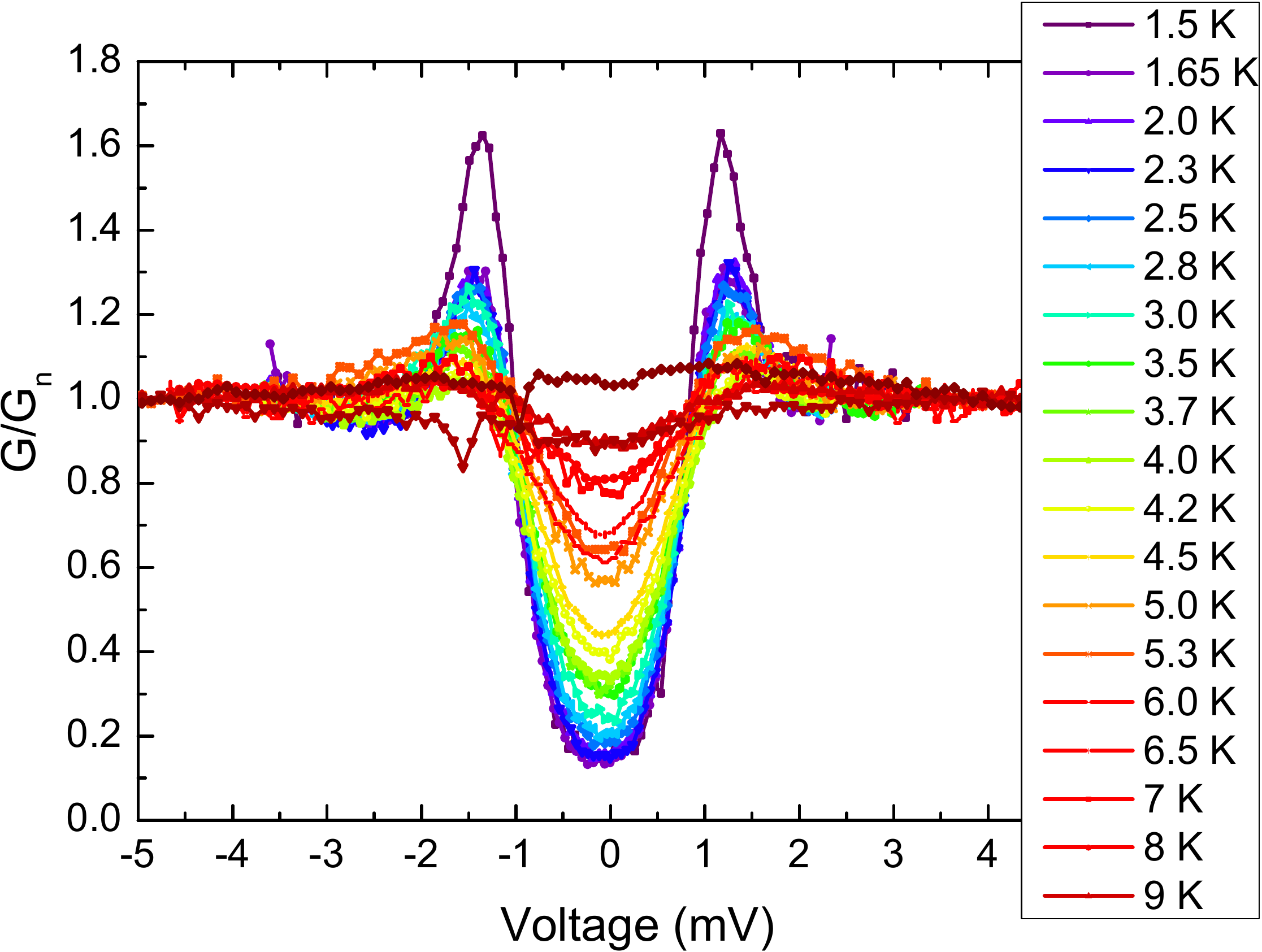}
\caption{Plot of normalized conductance $G/G_\text{n}$ vs. bias voltage at different temperatures, taken with a Nb tip on the Au(100) sample imaged in Fig.\ 3(b).}
\end{figure}

Figure 5 shows a scanning tunneling spectroscopy map taken with a Nb tip on a Bi$_2$Se$_3$ sample at 4.2 K. In addition to topography [Fig.\ 5(a)], a spectroscopy curve ($\text{d}I/\text{d}V$ versus $V$) was taken at each point in the image, with the bias voltage being swept from $-8.5$  to $6.5$ mV. Figure 5(b) shows a sample frame for $V = -8.5$ mV (see movie online). The data shows the tip is sensitive to spatial variations in the conductance. The singularity at the Nb gap edge enables the tip to resolve small spectroscopic features at extremely small bias voltages. This conductance map was taken over a 12 hour period, and at tunnel resistances as low as 5 M$\Omega$, indicating that the tips are stable. We stress the importance of tip stability (while remaining atomically sharp) at close proximity to the sample surface because at cryogenic temperatures it is the primary factor limiting an STM's ability to obtain spatially resolved conductance maps with fine energy resolution. 
\begin {figure}[t]
\centering
\includegraphics[width= 0.45\textwidth]{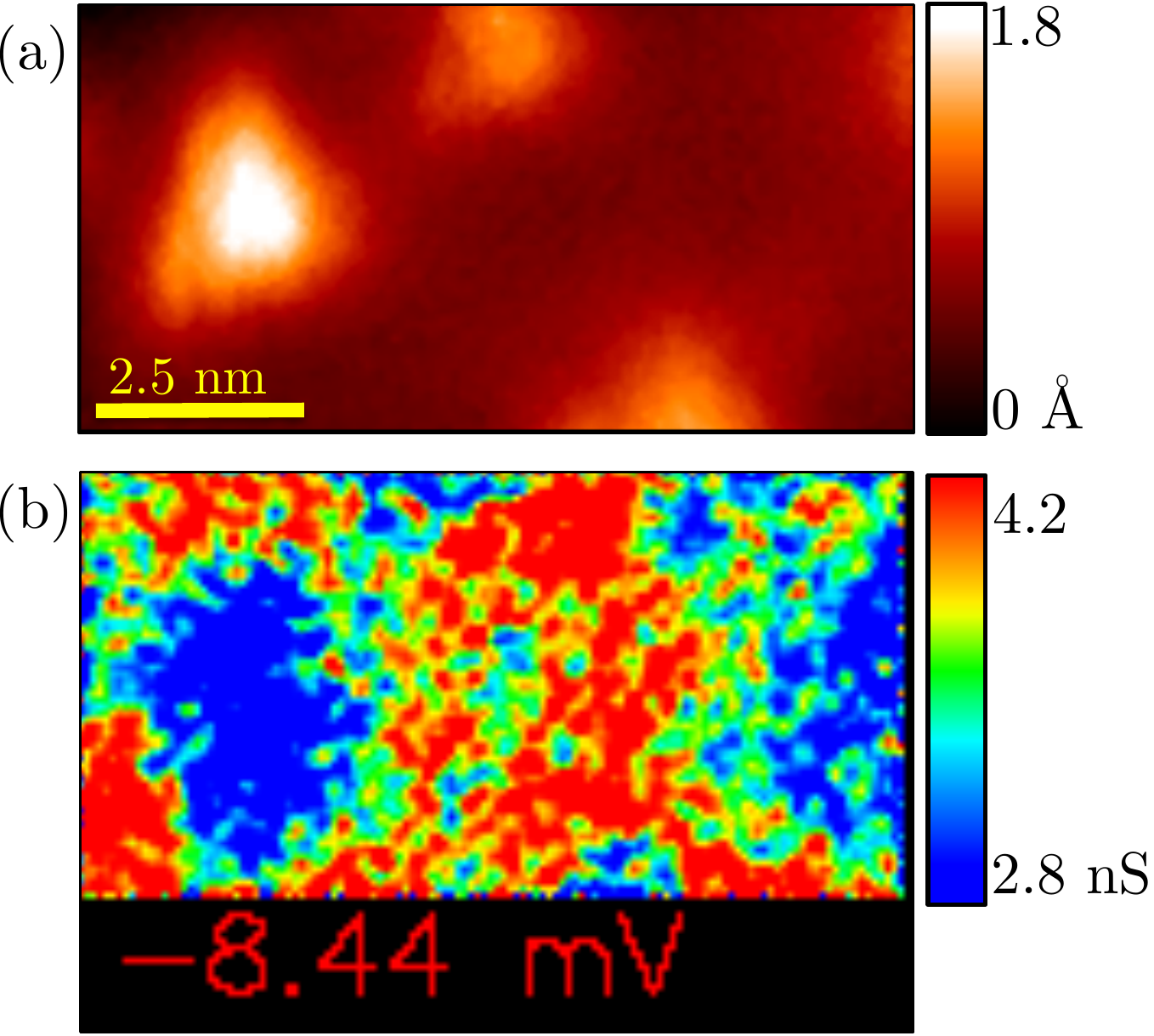}
\caption{Topographic and spectroscopic images taken with a Nb tip on a Bi$_2$Se$_3$ sample at 4.2 K. (a) Topography in the vicinity of triangular defects caused by the replacement of single Se atoms with Bi. (b) Conductance map (movie online). The movie shows slices of the conductance at fixed voltage as the voltage is swept from $-8.5$ mV to $6.5$ mV. }
\end{figure}

Finally, Fig.~6 shows a series of $\text{d}I/\text{d}V$ curves taken on a bulk Nb(100) sample at 30 mK at different tunneling resistances. The Nb tip and sample form an ultra-small Josephson junction. As expected, we see a small phase diffusive super-current at finite voltages.\cite{Naaman01} Examination of the plot reveals sub-gap features in the zero bias conductance peak of the phase diffusive supercurrent. These features are reproducible at different tunnel resistances, as well as with different tips at different locations on the sample. This behavior is consistent with $P(E)$ theory \cite{Averin86, Ingold94} which predicts peaks at voltages where incoherent Cooper pairs can radiate energy to electromagnetic modes of the environment. The reproducibility of such fine scale features is further evidence for the mechanical stability of the tip as well as a spectroscopic resolution of about 8 $\mu$V at ultra low temperatures.

\begin {figure}[t]
\centering
\includegraphics[width= 0.45\textwidth]{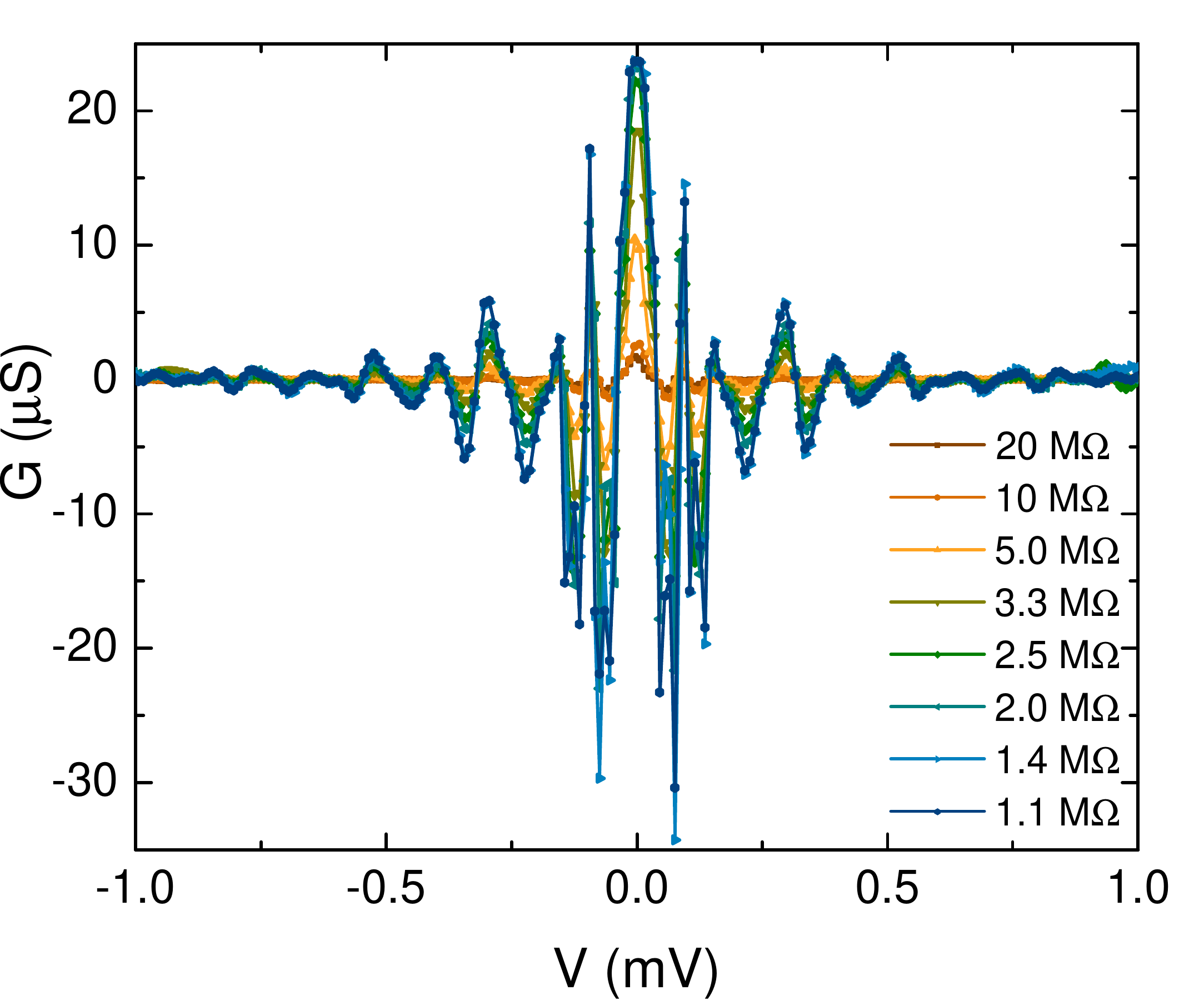}
\caption{Conductance ($\text{d}I/\text{d}V$) versus bias voltage $V$ data taken on a Nb(100) single crystal with a Nb tip at 30 mK for different tunnel resistances. The tip-to-sample bias voltage $V$ was set to 2 mV and the current was chosen to fix the tunnel resistance of the junction. The $z$-feedback loop was then turned off during the measurement, and the bias voltage $V$ was swept. The data shows the evolution of the zero bias conductance peak and additional sub-gap excitations with tunnel resistance.}
\end{figure}

In conclusion, we have described a reactive ion etching technique that requires minimal oversight to fabricate multiple sharp superconducting Nb tips. The performance of these tips was demonstrated via atomic resolution images, temperature dependent spectroscopy and a conductance map. Our results indicate that the tips are superconducting, mechanically stable and atomically sharp. Furthermore, the tips display excellent spectroscopic energy resolution at mK temperatures. Finally, we expect that our plasma-based etching technique may be extended to other materials for a range of applications. Our recipe yields multiple ultra-sharp probes which may be suitable in situations where the tip radius of curvature is an important factor, including field emission techniques, \cite{Zhu03, Givargizov93} atom manipulation techniques \cite{Loth12, Loth10} and STM based lithography. \cite{Schofield03, Haider09, Randall09, Pires10, Durrani13} 

The authors acknowledge many useful discussions on reactive ion etching techniques with G.~Porkolab and thank I.~Miotkowski and Y.~P.~Chen for the Bi$_2$Se$_3$ samples. Portions of this work were funded by the NSF under DMR-0605763.

\bibliography{bibliography}

\end{document}